\documentclass{Interspeech2024}
\interspeechcameraready 
\usepackage{amsmath,graphicx,multicol,multirow}

\title{Multi-Modal Automatic Prosody Annotation with \\ Contrastive Pretraining of Speech-Silence and Word-Punctuation}

\name[affiliation={1,2}]{Jinzuomu}{Zhong}
\name[affiliation={2}]{Yang}{Li}
\name[affiliation={2}]{Hui}{Huang}
\name[affiliation={1}]{Korin}{Richmond}
\name[affiliation={2}]{Jie}{Liu}
\name[affiliation={2}]{Zhiba}{Su}
\name[affiliation={2}]{Jing}{Guo}
\name[affiliation={2}]{Benlai}{Tang}
\name[affiliation={2}]{Fengjie}{Zhu}

\address{
 $^1$The Centre for Speech Technology Research, University of Edinburgh, United Kingdom \\
 $^2$Department of AI Technology, Transsion, China}
\email{j.zhong-12@sms.ed.ac.uk, korin.richmond@ed.ac.uk, benlai.tang@transsion.com}


\begin{document}

\maketitle

\begin{abstract}
In expressive and controllable Text-to-Speech (TTS), explicit prosodic features significantly improve the naturalness and controllability of synthesised speech. However, manual prosody annotation is labor-intensive and inconsistent. To address this issue, a two-stage automatic annotation pipeline is novelly proposed in this paper. In the first stage, we use contrastive pretraining of Speech-Silence and Word-Punctuation (SSWP) pairs to enhance prosodic information in latent representations. In the second stage, we build a multi-modal prosody annotator, comprising pretrained encoders, a text-speech fusing scheme, and a sequence classifier. Experiments on English prosodic boundaries demonstrate that our method achieves state-of-the-art (SOTA) performance with 0.72 and 0.93 f1 score for Prosodic Word and Prosodic Phrase boundary respectively, while bearing remarkable robustness to data scarcity.
\end{abstract}

\vspace{-1em}
\section{Introduction}


Recent advances in expressive Text-to-Speech (TTS) systems have made it possible to generate speech that is indistinguishable from human speech \cite{liu2021delightfultts, tan2024naturalspeech}. High naturalness of these systems is largely achieved by implicit modelling of prosody: 1) enriching text/phonetic representation with contextual information from text/phonetic pretraining such as \cite{tan2024naturalspeech, zhou2021enhancing, jia2021png, zhang2022mixed, li2023phoneme}; 2) self-supervised learning of prosody-related acoustic features such as pitch or other learned information from audio \cite{liu2021delightfultts, ren2020fastspeech, ren2022prosospeech, ye2023clapspeech, babianski2023granularity}; 3) enhancing model capabilities with generative modelling \cite{ren2021portaspeech, kim2021conditional}. However, implict modelling of prosody suffers controllability issues as it cannot correct prosody and pause errors when the model generates bad cases.

By contrast, conditioning the speech on explicit prosodic features such as hierarchical prosodic boundaries \cite{dai2022automatic, yuan2022low}, ToBI \cite{rosenberg2010autobi, zou21_interspeech}, and contrastive focus \cite{stephenson22_interspeech, sanders2023invert} explicitly models speech prosody, and thus offer two advantages: 1) improvement in naturalness by fine-grained prosodic information, and more importantly, 2) precise control over prosody and pause as demonstrated in these previous works. However, human annotation of prosodic features is time-consuming, expensive, and often inconsistent \cite{dai2022automatic} and therefore calls for automatic annotation.

Three categories of method are attempted to address automatic prosody annotation: 1) audio-only spectrogram analysis approach \cite{rosenberg2010autobi, suni2017hierarchical}, 2) text-only prosody prediction \cite{zou21_interspeech, stephenson22_interspeech, sloan19_ssw, chen2022character}, and 3) multi-modal prosody annotation \cite{dai2022automatic, yuan2022low, ananthakrishnan2007automatic, sridhar2008exploiting, qian2010automatic}. Among all these, multi-modal prosody annotation with pretrained text-speech model outperforms the others by utilizing inputs of both modalities, but the result is still far from satisfactory. Dai et al.\ \cite{dai2022automatic} use cross-attention to align and fuse acoustic information from a pretrained Conformer-ASR model \cite{gulati2020conformer} and text information from a pretrained BERT \cite{devlin2018bert} for Chinese prosodic boundary annotation. Yuan et al.\ \cite{yuan2022low} apply cross-lingual transfer learning of the same model for Mongolian. The major drawbacks of this approach are two-fold. 1) Implicit alignment between text and speech by cross-attention leads to high requirement for annotated data. 2) Phonetic posteriorgrams (PPGs) lack sufficient prosodic information \cite{li2021ppg} which is crucial for prosody annotation. These drawbacks motivate us to search for better representation with enriched prosody information that can improve the performance of prosody annotation. In this paper, we study the automatic annotation of English prosodic boundaries and its effect on synthesised speech.

Inspired by cross-modal contrastive learning, CLIP \cite{pmlr-v139-radford21a}, CLAP \cite{wu2023large, elizalde2023clap}, and its adaptation in the TTS community such as CLAPSpeech \cite{ye2023clapspeech}, we propose a two-stage training pipeline with novel pretraining strategy and simple model architecture that achieves SOTA performance on automatic annotation of English prosodic boundaries.
In the first stage, we propose the contrastive text-speech learning of a new unit, named Speech-Silence and Word-Punctuation (SSWP) pairs, as silence- and punctuation-related information is crucial for detecting prosodic boundary. Conformer \cite{guo2021recent} and pretrained BERT \cite{devlin2018bert} models are used as speech and text encoder respectively to extract the shared representation. In the second stage, i.e. prosody annotation, we add the text and audio latent representation of each SSWP and concatenate text-speech latent representations of all SSWPs in a sentence to represent speech prosody. Both modalities of inputs are used due to the variability of speech prosody given the same text. Finally, a standard bi-LSTM sequence classification network maps the combined text-speech latent representations to their corresponding prosodic boundary classes.

To summarize, our approach mainly contributes to the following three areas:
\begin{itemize}
    \item To the best of our knowledge, we are the first to introduce contrastive learning to prosody annotation. We propose a novel pretraining of Speech-Silence and Word-Punctuation (SSWP) pairs to enhance the prosodic information of the learned representation.
    \item We propose a novel multi-modal prosody annotation architecture in the second stage, comprising pretrained encoders, straightforward yet effective text-speech feature fusion, and sequence classification network.
    \item We achieve SOTA performance on English posodic boundary annotation task with higher resilience against limited data.
\end{itemize}

\begin{figure*}[!htbp]
\centering
\includegraphics[width=1\linewidth, height=0.5\linewidth, trim = 25 170 75 28, clip]{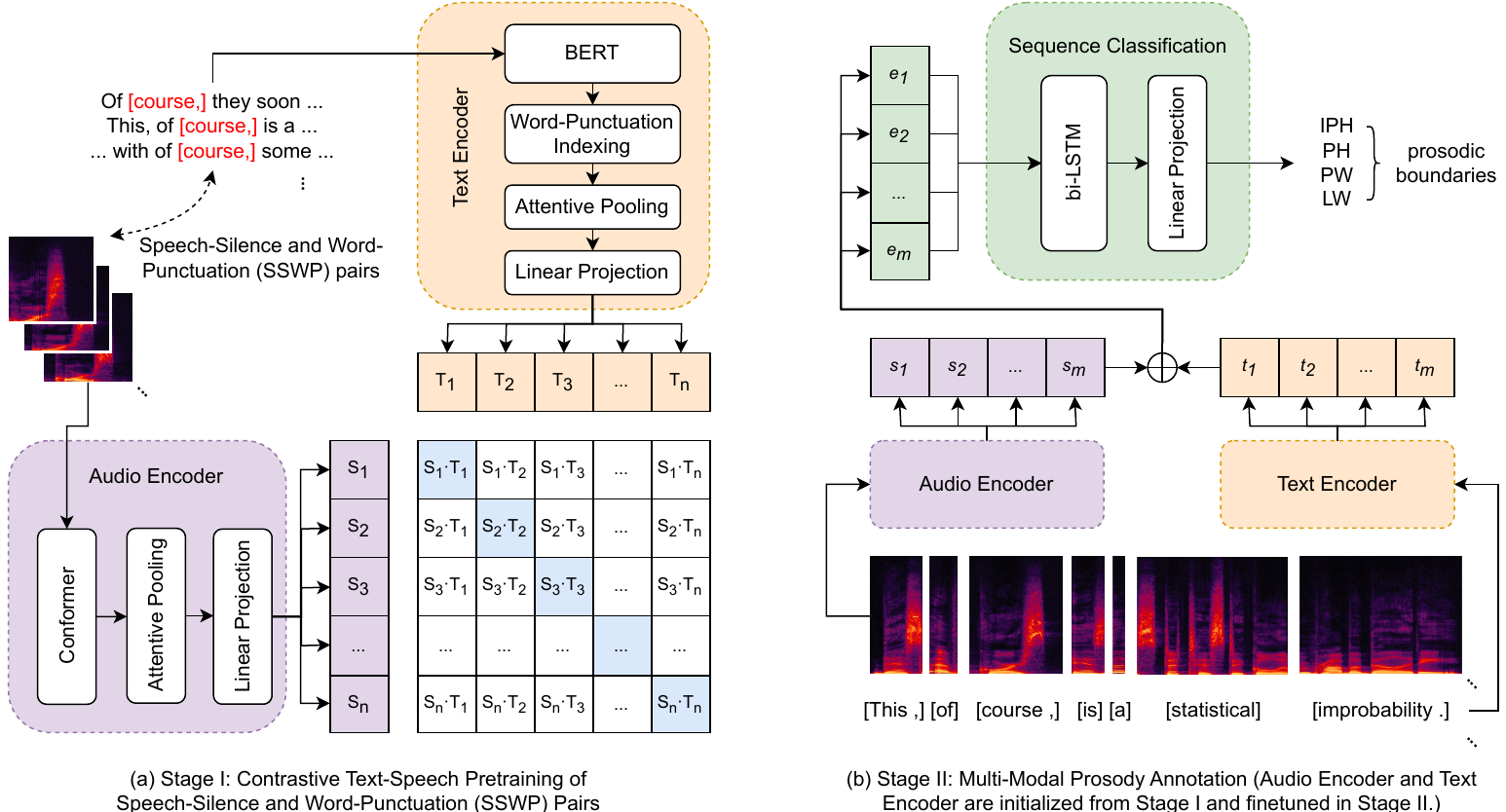}
\caption{The architecture of our proposed two-stage training pipeline.}
\vspace{-1.5em}
\label{fig:model}
\end{figure*}

\vspace{-1em}
\section{Method}
As shown in Figure \ref{fig:model}, our proposed method automatically annotates prosodic boundaries via a two-stage training pipeline. In Section \ref{ssec:prosodic_boundary}, we introduce the standard of English prosodic boundaries adopted in this work and how these features are annotated by human, a process that our approach aim to mimick. In Sections \ref{ssec:pretraining} and \ref{ssec:downstream}, we introduce the first stage contrastive text-speech pretraining of SSWP pairs and the second stage multi-modal prosody annotation, respectively.
\ifinterspeechfinal
    Code is openly available at \url{https://github.com/jzmzhong/Automatic-Prosody-Annotator-with-SSWP-CLAP/}
\else
    Code is openly available at [URL redacted for author anonymity].
\fi

\subsection{Prosodic Boundary Annotation}
\label{ssec:prosodic_boundary}

The prosody annotation adopted in this work categorizes the prosodic boundaries of English speech into four levels, including Lexicon Word (LW), Prosodic Word (PW), Prosodic Phrase (PPH), and Intonational Phrase (IPH) \cite{ selkirk1980prosodic},  as shown in Table \ref{tab:ling}. Human annotators label the boundaries based on both text (e.g. syntax, semantics) and speech (e.g. pitch, pace, silence).

\begin{table}[h!]
\vspace{-0.5em}
\caption{Four-level prosodic boundaries of English speech.}
\label{tab:ling}
\vspace{-0.75em}
\centering
\setlength\tabcolsep{2.1pt}
\begin{tabular}{|l|ccccccccccc|}
\hline
IPH &
 \multicolumn{9}{c|}{We must urge representatives to push for reforms.} \\
\hline
PPH &
 \multicolumn{4}{c|}{We must urge representatives} &
 \multicolumn{5}{c|}{to push for reforms.} \\
\hline
PW &
 \multicolumn{2}{c|}{We must} &
 \multicolumn{2}{c|}{urge representatives} &
 \multicolumn{5}{c|}{to push for reforms.} \\
\hline
LW &
 \multicolumn{1}{c|}{We} &
 \multicolumn{1}{c|}{must} &
 \multicolumn{1}{c|}{urge} &
 \multicolumn{1}{c|}{representatives} &
 \multicolumn{1}{c|}{to} &
 \multicolumn{1}{c|}{push} &
 \multicolumn{1}{c|}{for} &
 \multicolumn{1}{c|}{reforms} &
 \multicolumn{1}{c|}{.} \\
\hline
\end{tabular}
\vspace{-1.5em}
\end{table}

\subsection{Contrastive Text-Speech Pretraining of SSWP Pairs}
\label{ssec:pretraining}

The overall architecture of this first stage is shown in Figure \ref{fig:model}(a). The pretraining aims to extract the prosodic space by contrasting text-speech pairs of the same phonetic content under different contexts. 

\noindent\textbf{Speech-Silence \& Word-Punctuation (SSWP) Pairs}
Prosodic boundaries have a high correlation with both punctuations in text and silence segments in speech as they usually indicate the possible existence of PW, PPH, or IPH boundaries \cite{steinhauer2003electrophysiological}.
Neither prompts/sentences \cite{wu2023large, elizalde2023clap} nor words \cite{dai2022automatic} are ideal contrastive units for pretraining as they do not include boundary-related information. We propose to contrastively pretrain Speech-Silence \& Word-Punctuation (SSWP) pairs, where words with following punctuations are paired with speech segments with following silence segments if there are any.

\noindent\textbf{Text \& Audio Encoder}
We choose Conformer \cite{gulati2020conformer} as the audio encoder and BERT \cite{devlin2018bert} as the text encoder due to its wide application in various speech and text tasks, similar to the encoder choices made in \cite{ye2023clapspeech, wu2023large, elizalde2023clap}. Speech-silence segments are passed into Conformer blocks. An attentive pooling layer is used to handle variable length and a linear projection layer is used to reshape latent representations into the dimension of the joint prosodic space, following \cite{ye2023clapspeech, pmlr-v139-radford21a, wu2023large, elizalde2023clap}. The entire text utterances of text-punctuation segments are passed into BERT to include contextual information. A following indexing layer then selects the corresponding subwords for the SSWP in a sentence, following \cite{ye2023clapspeech}. Similar to the audio encoder, an attentive pooling layer and a linear projection layer map the text representation to the joint prosodic space. Let the speech-silence segment of index $i$ be $X_{SS}^{(i)}$, the text utterance that contains the corresponding word-punctuation be $X_{WP}^{(i)}$, and the subword start \& end indexes of the word-punctuation in the utterance be $(i_m, i_n)$. The text and speech representation of each SSWP, $T_{i}$ and $S_{i}$ can be written as follows.
\vspace{-0.5em}
\begin{equation}
\begin{split}
& T_{i}=\text{Linear}(\text{Pool}(\text{BERT}(X_{WP}^{(i)}){[i_m:i_n]}))
\end{split}
\end{equation}
\vspace{-1.5em}
\begin{equation}
\begin{split}
& S_{i}=\text{Linear}(\text{Pool}(\text{Conformer}(X_{SS}^{(i)})))
\end{split}
\end{equation}
\noindent\textbf{Contrastive Pretraining}
The text-speech model is trained with the contrastive learning paradigm between the text and speech embeddings of paired extended-words, $T_{i}$ and $S_{i}$, following the same loss function as in \cite{ye2023clapspeech, pmlr-v139-radford21a, wu2023large, elizalde2023clap}:
\vspace{-0.5em}
\begin{equation}
\begin{split}
\mathcal{L_\text{pretrain}} = \frac{1}{2n} \sum_{i=1}^{n} & {(\log\frac{exp(S_{i} \cdot T_{i} / \tau)}{\sum_{j=1}^{n} exp(S_{i} \cdot T_{j} / \tau)}} \\
& {+\log\frac{exp(T_{i} \cdot S_{i} / \tau)}{\sum_{j=1}^{n} exp(T_{i} \cdot S_{j} / \tau)})}
\end{split}
\end{equation}
\noindent where $n$ is the number of samples in a batch, and $\tau$ is a learnable temperature parameter for scaling the loss. The negative examples $T_{j}, S_{j}$ where $j \neq i$, are sampled randomly from a group of SSWP pairs with the same phonetic contents but different contexts.

\subsection{Multi-modal Prosody Annotation}
\label{ssec:downstream}

Since the same text can be expressed in different speech prosody dependant on the speaker, multi-modal information is used. The overall architecture of this second stage is shown in Figure \ref{fig:model}(b). SSWP pairs in a sentence are passed into the Text \& Audio Encoder pretrained from the first stage. The text and speech representations of a sentence, $t_{1,2,...m}$ and $s_{1,2,...m}$ are then simply added to obtain a latent representation with enhanced prosodic information $e_{1,2,...m}$.
A standard bi-LSTM followed by a linear projection layer with softmax function is used to form the sequence classification network that predicts the probability distribution of prosodic boundaries $\hat y_{1,2,...m}$, written as follows:
\vspace{-0.5em}
\begin{equation}
\hat y_{1,2,...m} = \sigma(\text{Linear}(\text{biLSTM}(t_{1,2,...m} + s_{1,2,...m})))
\vspace{-0.5em}
\end{equation}
\noindent where $\sigma$ denotes softmax activation. We adopt Cross Entropy Loss as the training objective of the classification model.

\vspace{-0.5em}
\section{Experiments}
\label{sec:exp}

\subsection{Datasets}
\label{ssec:datasets}

\noindent\textbf{Pretraining}
An open-source English ASR corpus, LibriSpeech \cite{panayotov2015librispeech}, is used for the first stage pretraining. 976.6 hours containing all subsets except for \texttt{test-clean} are used for training, and the remaining 5.4 hours are used for validation. Punctuation is inserted by using Whisper \cite{radford2023robust}. Word boundaries are obtained using Montreal Forced Aligner \cite{mcauliffe2017montreal}.

\noindent\textbf{Prosody Annotation}
A proprietary English TTS corpus with prosodic boundary annotation is used for the second stage downstream prosody annotation due to the lack of high-quality English open-source corpus with prosodic boundary annotation. Performance of different systems in three scenarios are evaluated: on low/medium resource and on unseen/seen speakers, listed in Table \ref{tab:data}. In the unseen speaker scenario, the trained model is used to predict prosodic boundaries of out-of-domain speakers; whereas in the seen speaker scenario, the trained model is used to predict prosodic boundaries of out-of-domain utterances by an in-domain speaker. Low resource unseen speaker is not evaluated since: i) zero-shot unseen speaker with limited one-speaker data is highly contingent upon the similarity in prosody between the unseen speaker and the trained speaker, ii) such a scenario is hardly encountered in the development of TTS systems. To validate/test the seen speaker scenario, we must select out-of-domain utterances from the same in-domain speaker corpus for training; due to limited data, an unavoidably small valid/test data size of 500 utterances is adopted.

\begin{table}[h!]
\vspace{-0.5em}
\captionsetup{justification=centering}
\caption{Data composition of three annotation scenarios. \\
Res. - Resource; Tar. - Target; Spk. - Speaker. \\
Both the number of utterances and the number of speakers are listed, e.g. 21k (7spk) denotes 21,000 utterances by 7 speakers.}
\label{tab:data}
\vspace{-0.75em}
\centering
\renewcommand{\arraystretch}{1}
\setlength\tabcolsep{3.5pt}
\begin{tabular}{cc|ccc}
\hline
Res. Scale & Tar. Spk. & Train & Valid & Test \\ \hline
Medium & Unseen & 21k (7Spk) & 2k (1Spk) & 2k (1Spk) \\
Medium & Seen & 21k (7Spk) & 500 (1Spk) & 500 (1Spk) \\
Low & Seen & 8k (1Spk) & 500 (1Spk) & 500 (1Spk) \\ \hline
\end{tabular}
\vspace{-1.5em}
\end{table}

\subsection{Baselines}
\label{ssec:baselines}

We adopt two baselines: text-only prediction baseline named \texttt{Text Baseline}, and multi-modal annotation baseline named \texttt{Multi-modal Baseline}. For the \texttt{Text Baseline}, we use only text as input to a pretrained BERT\footnote{\label{note1}https://huggingface.co/bert-base-uncased} followed by a fully-connected layer as the classifier. For the \texttt{Multi-modal Baseline}, we reproduce the work in \cite{dai2022automatic} as faithfully as possible and reapply the model on the same English dataset - both text and speech are passed into respective encoder and then combined using attention followed by a Transformer decoder and a fully-connected layer as classifier. We use the same aforementioned BERT as text encoder and a pretrained Conformer-ASR model\footnote{\label{note2}https://huggingface.co/nvidia/stt\_en\_conformer\_ctc\_large} as audio encoder. To handle the mismatch in length between the latent representation of subword units in BERT and the prosodic labels of word units in classification targets, we adopt average pooling to merge the subword latent representations of each word. This is the only difference we have made in reproducing \texttt{Text Baseline} and \texttt{Multi-modal Baseline} and we do not think such adaption due to different experiment language would be an issue. We achieve comparable baseline results to the original paper, as shown in Table \ref{tab:result}.

\subsection{Training Configurations}
\label{ssec:configs}

\noindent\textbf{Model Configurations}
The Conformer and the pretrained BERT that we adopt for the audio and text encoders of the \texttt{Proposed Work} are much smaller in size compared to those of the two baselines mentioned above. To save computing resources and speed up training, we use fewer parameters in our \texttt{Proposed Work} but still use openly available large pretrained models as baselines for fair and benevolent comparison. We use 12-layer BERT of 768 dimension size\footnotemark[1] and 18-layer Conformer of 512 dimension size\footnotemark[2] for the two baselines. We use 4-layer BERT of 256 dimension size\footnote{https://huggingface.co/prajjwal1/bert-mini} and 4 layers Conformer of 256 dimension size for our proposed work.

\noindent\textbf{Pretraining}
We pretrain the text-speech model on 4 Nvidia 3090Ti GPUs with a batch size of 2,048 text-speech pairs (512 pairs per GPU). We use the Adam optimizer with an initial learning rate of 1e-4. We train the text-speech model for 50 epochs with a cosine learning rate schedule.

\noindent\textbf{Prosody Annotation} 
We train the classification model on 1 Nvidia 3090Ti GPU with a batch size of 16 sentences. We use the Adam optimizer with an initial learning rate of 1e-5. We train the classification model for 50 epochs with a cosine learning rate schedule.

\begin{table*}[ht]
\centering
\setlength\tabcolsep{3.0pt}
\renewcommand{\arraystretch}{1}
\captionsetup{justification=centering}
\caption{Results of previous benchmarks and the proposed work. Res. - Resource, Spk. - Speaker, prec - precision, rec - recall.}
\label{tab:result}
\vspace{-1em}
\begin{tabular}{l|cccccc|cccccc|cccccc}
\hline
\multicolumn{1}{c|}{\multirow{3}{*}{Systems}} &
  \multicolumn{6}{c|}{Medium Res. Unseen Spk.} &
  \multicolumn{6}{c|}{Medium Res. Seen Spk.} &
  \multicolumn{6}{c}{Low Res. Seen Spk.} \\ \cline{2-19} 
\multicolumn{1}{c|}{} &
  \multicolumn{3}{c|}{PW} &
  \multicolumn{3}{c|}{PPH} &
  \multicolumn{3}{c|}{PW} &
  \multicolumn{3}{c|}{PPH} &
  \multicolumn{3}{c|}{PW} &
  \multicolumn{3}{c}{PPH} \\ \cline{2-19} 
\multicolumn{1}{c|}{} &
  prec &
  rec &
  \multicolumn{1}{l|}{f1} &
  \multicolumn{1}{c}{prec} &
  rec &
  f1 &
  prec &
  rec &
  \multicolumn{1}{l|}{f1} &
  \multicolumn{1}{c}{prec} &
  rec &
  f1 &
  prec &
  rec &
  \multicolumn{1}{l|}{f1} &
  \multicolumn{1}{c}{prec} &
  rec &
  f1 \\ \hline
Text Baseline &
  \multicolumn{1}{r}{0.35} &
  0.56 &
  0.43 &
  0.88 &
  0.73 &
  0.80 &
  \multicolumn{1}{r}{0.38} &
  0.46 &
  0.42 &
  0.88 &
  0.63 &
  0.74 &
  \multicolumn{1}{r}{0.35} &
  0.59 &
  0.44 &
  0.87 &
  0.65 &
  0.75 \\ \hline
Multi-modal Baseline &
  \multicolumn{1}{r}{0.44} &
  0.48 &
  0.46 &
  0.84 &
  0.83 &
  0.84 &
  \multicolumn{1}{r}{0.50} &
  0.43 &
  0.46 &
  0.86 &
  0.75 &
  0.80 &
  \multicolumn{1}{r}{0.35} &
  0.59 &
  0.44 &
  0.87 &
  0.63 &
  0.73 \\ \hline
Proposed Work &
  \multicolumn{1}{r}{\textbf{0.76}} &
  \textbf{0.58} &
  \textbf{0.66} &
  \textbf{0.94} &
  \textbf{0.93} &
  \textbf{0.93} &
  \multicolumn{1}{r}{\textbf{0.70}} &
  \textbf{0.74} &
  \textbf{0.72} &
  \textbf{0.91} &
  \textbf{0.93} &
  \textbf{0.92} &
  \multicolumn{1}{r}{\textbf{0.70}} &
  \textbf{0.75} &
  \textbf{0.73} &
  \textbf{0.93} &
  \textbf{0.89} &
  \textbf{0.91} \\ \hline
\end{tabular}
\vspace{-0.75em}
\end{table*}

\begin{table*}[ht]
\centering
\setlength\tabcolsep{3.0pt}
\captionsetup{justification=centering}
\caption{Results of ablation studies. Only one component is ablated at each time to better investigate its efficacy.}
\label{tab:ablation}
\vspace{-1em}
\begin{tabular}{l|cccccc|cccccc|cccccc}
\hline
\multicolumn{1}{c|}{\multirow{3}{*}{Systems}} &
  \multicolumn{6}{c|}{Medium Res. Unseen Spk.} &
  \multicolumn{6}{c|}{Medium Res. Seen Spk.} &
  \multicolumn{6}{c}{Low Res. Seen Spk.} \\ \cline{2-19} 
\multicolumn{1}{c|}{} &
  \multicolumn{3}{c|}{PW} &
  \multicolumn{3}{c|}{PPH} &
  \multicolumn{3}{c|}{PW} &
  \multicolumn{3}{c|}{PPH} &
  \multicolumn{3}{c|}{PW} &
  \multicolumn{3}{c}{PPH} \\ \cline{2-19} 
\multicolumn{1}{c|}{} &
  \multicolumn{1}{c}{prec} &
  rec &
  \multicolumn{1}{l|}{f1} &
  \multicolumn{1}{c}{prec} &
  rec &
  f1 &
  \multicolumn{1}{c}{prec} &
  rec &
  \multicolumn{1}{l|}{f1} &
  \multicolumn{1}{c}{prec} &
  rec &
  f1 &
  \multicolumn{1}{c}{prec} &
  rec &
  \multicolumn{1}{l|}{f1} &
  \multicolumn{1}{c}{prec} &
  rec &
  f1 \\ \hline
Proposed Work &
  \textbf{0.76} &
  0.58 &
  \textbf{0.66} &
  \textbf{0.94} &
  0.93 &
  \textbf{0.93} &
  \textbf{0.70} &
  \textbf{0.74} &
  \textbf{0.72} &
  \textbf{0.91} &
  0.93 &
  \textbf{0.92} &
  \textbf{0.70} &
  \textbf{0.75} &
  \textbf{0.72} &
  0.93 &
  0.89 &
  0.91 \\ \hline
w/o Contrastive Pretrain &
  0.74 &
  0.48 &
  0.58 &
  0.89 &
  \textbf{0.95} &
  0.92 &
  0.68 &
  0.68 &
  0.68 &
  0.88 &
  0.93 &
  0.91 &
  0.66 &
  \textbf{0.75} &
  0.70 &
  \textbf{0.94} &
  0.89 &
  \textbf{0.92} \\ \hline
w/o Any Pretrain &
  0.70 &
  0.44 &
  0.54 &
  0.88 &
  \textbf{0.95} &
  0.91 &
  0.64 &
  \textbf{0.74} &
  0.69 &
  0.89 &
  0.94 &
  \textbf{0.92} &
  0.64 &
  0.73 &
  0.68 &
  0.91 &
  \textbf{0.90} &
  0.91 \\ \hline
w/o SSWP &
  0.64 &
  0.48 &
  0.55 &
  0.87 &
  0.84 &
  0.85 &
  0.64 &
  0.69 &
  0.66 &
  \textbf{0.91} &
  0.74 &
  0.81 &
  0.57 &
  0.65 &
  0.61 &
  0.86 &
  0.75 &
  0.80 \\ \hline
w/o bi-LSTM &
  \multicolumn{1}{l}{0.65} &
  \textbf{0.60} &
  0.62 &
  \multicolumn{1}{l}{0.92} &
  0.91 &
  0.91 &
  \multicolumn{1}{l}{\textbf{0.70}} &
  0.58 &
  0.64 &
  \multicolumn{1}{l}{\textbf{0.91}} &
  \textbf{0.95} &
  0.88 &
  \multicolumn{1}{l}{0.55} &
  0.61 &
  0.58 &
  \multicolumn{1}{l}{0.93} &
  0.80 &
  0.86 \\ \hline
\end{tabular}
\vspace{-1.5em}
\end{table*}

\vspace{-1em}
\section{Results}
\label{sec:results}

\subsection{Objective Evaluation}
\label{ssec:objective}

The results of our proposed work, compared with previous benchmarks, are shown in Table \ref{tab:result}. Due to limited space, the results of IPH boundary are not listed since they are all 0.99-1.00 in terms of f1 score. The results of LW boundary is not listed either as it is the default tag with an f1 score of 0.97-0.98. Different prosodic boundary annotators mainly differ in PW and PPH boundaries detection.

Our \texttt{Proposed Work} improves the \texttt{Multi-modal Baseline}, by \textbf{0.20-0.29} and \textbf{0.11-0.18} f1 score of PW and PPH boundaries respectively across all three annotation scenarios. Additionally, our proposed work achieves near-human expert annotation accuracy on the f1 score of PPH boundaries, ranging between 0.91-0.93. These results demonstrate both higher performance and broader applicability of our proposed work. Moreover, our proposed work has higher resilience to data scarcity. In low resource seen speaker scenario, the efficacy of the multi-modal baseline vanishes compared with text baseline, while our proposed work can achieve remarkable 0.29 and 0.18 f1 score gain for PW and PPH boundaries respectively. Objective results prove that our proposed work, despite being smaller in model size, gives better performance, broader applicability, and higher resilience to data scarcity.

\vspace{-0.5em}
\subsection{Ablation Studies}
\label{ssec:ablation}

We conduct extensive ablation studies to show the efficacy of each module in our proposed work, as shown in Table \ref{tab:ablation}. The best system \texttt{Proposed Work}, as shown on the first row, outperforms all ablated systems on most objective evaluation metrics, proving the necessity of each aspect of our proposed approach. \texttt{w/o Contrastive Pretrain} system does not use the pretrained text-speech model from the first stage to initialize classification model weights in the second stage, but the text encoder is still initialized by weights from the pretrained BERT. \texttt{w/o Any Pretrain} system does not use any pretrained model to initialize classification model weights. \texttt{w/o SSWP} system uses tokens instead of SSWP as text-speech pairs both in the first stage and second stage. \texttt{w/o bi-LSTM} system removes the bi-LSTM network from the best system and relies on one linear projection layer to classify prosodic boundaries.

Contrastive pretraining, SSWP, and sequence classification are found to be significantly contributing factors, as evidenced by the comparison between the best system and the ablated systems: 

1) Contrastive pretraining contributes to a 0.08 increase, compared with BERT pretraining alone, and a 0.12 increase, compared with no pretraining, in PW boundary f1 score under the medium resource unseen speaker scenario. This is due to the richer prosody representation extracted from contrastive pretraining in the first stage, compared with pure BERT or unpretrained representations. Under the medium resource seen speaker scenario, the effect of contrastive pretraining drops to a 0.04 increase in PW boundary f1 score, compared with purely BERT pretraining. This shows that contrastive pretraining is more effective in unseen speaker scenario due to its generalization capability learned from large pretrained data.

2) SSWP contributes to a 0.06-0.11 increase in PW f1 score and 0.08-0.11 PPH f1 score under all three scenarios. SSWP is the only component that achieves significant gains on PPH boundary classification. This is because SSWP, as a linguistically motivated design, is a more appropriate text-speech pair for prosody representation and prosody annotation tasks.

3) The bi-LSTM classification network is useful, especially in low resource scenario, contributing to a 0.14 increase in PW boundary f1 score, and a 0.05 in PPH boundary f1 score. The contextual information included in the recurrent network makes the overall system more robust to smaller amounts of data.

\vspace{-0.5em}
\subsection{Subjective Evaluation}
\label{ssec:subjective}

We also conduct an AB Preference evaluation on an open-source corpus of an unseen speaker, LJSpeech \cite{ljspeech17}, using the \texttt{Multi-modal Baseline} and \texttt{Proposed Work} to annotate the prosodic boundaries of the entire corpus automatically.
Both systems are trained using Conformer-FastSpeech2 \cite{guo2021recent} and Hifi-GAN \cite{kong2020hifi}, with a 16-dimension embedding encoding the prosodic boundary labels concatenated to the phone embedding as input. All models are trained for 200k steps with a batch size of 32.
30 native speakers are asked to compare the 30 utterances from each system and choose the one they prefer.
The results are shown in Table \ref{tab:mos}. Our proposed work achieves a \textbf{7.58\%} improvement in AB Preference compared with the previous benchmark. This demonstrates slight improvement in naturalness of the synthesised audio by our approach, in addition to the higher annotation accuracy for developing controllable TTS systems demonstrated in previous sections.
\ifinterspeechfinal
    Demos are openly available at \url{https://jzmzhong.github.io/Automatic-Prosody-Annotator-With-SSWP-CLAP/}
\else
    Demos are openly available at [URL redacted for author anonymity].
\fi

\begin{table}[h!]
\vspace{-0.5em}
\captionsetup{justification=centering}
\caption{AB Preference results with 95\% confidence interval calculated for TTS systems trained on a corpus annotated by different prosody annotators. (p-value = 0.033)}
\label{tab:mos}
\vspace{-0.75em}
\centering
\renewcommand{\arraystretch}{1}
\begin{tabular}{l|c}
\hline
Prosody Annotation System & AB Preference (\%) \\ \hline
Multi-modal Baseline & 46.21 ± 3.33 \\
Proposed Work & 53.79 ± 3.33 \\ \hline
\end{tabular}
\vspace{-1em}
\end{table}

\vspace{-1em}
\section{Conclusions \& Future Work}
\label{sec:conclusions}

In this paper, we propose a novel two-stage pipeline that achieves SOTA performance on automatic annotation of English prosodic boundaries. With the design of contrastive text-speech pretraining of SSWP pairs, text and audio encoders learn richer prosody information. We also use aligned feature fusion and a sequence classification network to improve prosody annotation with contextual information.
In the future, we will investigate the proposed work in cross-lingual scenarios and on other prosodic features. We will also investigate the possibility of developing a unified TTS annotation tool covering phonetic information (out-of-vocabulary words, heteronyms, etc.) and other speech information (emotion, style, accent, etc.).


\newpage

\bibliographystyle{IEEEtran}
\bibliography{mybib}

\begin{thebibliography}{10}
\providecommand{\url}[1]{#1}
\csname url@samestyle\endcsname
\providecommand{\newblock}{\relax}
\providecommand{\bibinfo}[2]{#2}
\providecommand{\BIBentrySTDinterwordspacing}{\spaceskip=0pt\relax}
\providecommand{\BIBentryALTinterwordstretchfactor}{4}
\providecommand{\BIBentryALTinterwordspacing}{\spaceskip=\fontdimen2\font plus
\BIBentryALTinterwordstretchfactor\fontdimen3\font minus \fontdimen4\font\relax}
\providecommand{\BIBforeignlanguage}[2]{{%
\expandafter\ifx\csname l@#1\endcsname\relax
\typeout{** WARNING: IEEEtran.bst: No hyphenation pattern has been}%
\typeout{** loaded for the language `#1'. Using the pattern for}%
\typeout{** the default language instead.}%
\else
\language=\csname l@#1\endcsname
\fi
#2}}
\providecommand{\BIBdecl}{\relax}
\BIBdecl

\bibitem{liu2021delightfultts}
Y.~Liu, Z.~Xu, G.~Wang, K.~Chen, B.~Li, X.~Tan, J.~Li, L.~He, and S.~Zhao, ``{DelightfulTTS: The Microsoft Speech Synthesis System for Blizzard Challenge 2021},'' \emph{arXiv preprint arXiv:2110.12612}, 2021.

\bibitem{tan2024naturalspeech}
X.~Tan, J.~Chen, H.~Liu, J.~Cong, C.~Zhang, Y.~Liu, X.~Wang, Y.~Leng, Y.~Yi, L.~He \emph{et~al.}, ``{NaturalSpeech: End-to-End Text to Speech Synthesis with Human-Level Quality},'' \emph{IEEE Transactions on Pattern Analysis and Machine Intelligence}, 2024.

\bibitem{zhou2021enhancing}
Y.~Zhou, C.~Song, J.~Li, Z.~Wu, Y.~Bian, D.~Su, and H.~Meng, ``{Enhancing Word-Level Semantic Representation via Dependency Structure for Expressive Text-to-Speech Synthesis},'' in \emph{Proc. Interspeech}, 2022, pp. 5518--5522.

\bibitem{jia2021png}
Y.~Jia, H.~Zen, J.~Shen, Y.~Zhang, and Y.~Wu, ``{PnG BERT: Augmented BERT on Phonemes and Graphemes for Neural TTS},'' in \emph{Proc. Interspeech}, 2021, pp. 151--155.

\bibitem{zhang2022mixed}
G.~Zhang, K.~Song, X.~Tan, D.~Tan, Y.~Yan, Y.~Liu, G.~Wang, W.~Zhou, T.~Qin, T.~Lee, and S.~Zhao, ``{Mixed-Phoneme BERT: Improving BERT with Mixed Phoneme and Sup-Phoneme Representations for Text to Speech},'' in \emph{Proc. Interspeech}, 2022, pp. 456--460.

\bibitem{li2023phoneme}
Y.~A. Li, C.~Han, X.~Jiang, and N.~Mesgarani, ``{Phoneme-Level Bert for Enhanced Prosody of Text-To-Speech with Grapheme Predictions},'' in \emph{Proc. ICASSP}, 2023, pp. 1--5.

\bibitem{ren2020fastspeech}
Y.~Ren, C.~Hu, X.~Tan, T.~Qin, S.~Zhao, Z.~Zhao, and T.-Y. Liu, ``{FastSpeech 2: Fast and High-Quality End-to-End Text to Speech},'' in \emph{Proc. ICLR}, 2020.

\bibitem{ren2022prosospeech}
Y.~Ren, M.~Lei, Z.~Huang, S.~Zhang, Q.~Chen, Z.~Yan, and Z.~Zhao, ``Prosospeech: Enhancing prosody with quantized vector pre-training in text-to-speech,'' in \emph{Proc. ICASSP}, 2022, pp. 7577--7581.

\bibitem{ye2023clapspeech}
Z.~Ye, R.~Huang, Y.~Ren, Z.~Jiang, J.~Liu, J.~He, X.~Yin, and Z.~Zhao, ``{CLAPSpeech: Learning Prosody from Text Context with Contrastive Language-Audio Pre-Training},'' in \emph{Proc. ACL}, 2023, pp. 9317--9331.

\bibitem{babianski2023granularity}
M.~Babia{\'n}ski, K.~Pokora, R.~Shah, R.~Sienkiewicz, D.~Korzekwa, and V.~Klimkov, ``{On Granularity of Prosodic Representations in Expressive Text-to-Speech},'' in \emph{2022 IEEE Spoken Language Technology Workshop (SLT)}.\hskip 1em plus 0.5em minus 0.4em\relax IEEE, 2023, pp. 892--899.

\bibitem{ren2021portaspeech}
Y.~Ren, J.~Liu, and Z.~Zhao, ``{PortaSpeech: Portable and High-Quality Generative Text-to-Speech},'' \emph{Proc. NeurIPS}, vol.~34, pp. 13\,963--13\,974, 2021.

\bibitem{kim2021conditional}
J.~Kim, J.~Kong, and J.~Son, ``{Conditional Variational Autoencoder with Adversarial Learning for End-to-End Text-to-Speech},'' in \emph{International Conference on Machine Learning}.\hskip 1em plus 0.5em minus 0.4em\relax PMLR, 2021, pp. 5530--5540.

\bibitem{dai2022automatic}
Z.~Dai, J.~Yu, Y.~Wang, N.~Chen, Y.~Bian, G.~Li, D.~Cai, and D.~Yu, ``{Automatic Prosody Annotation with Pre-Trained Text-Speech Model},'' in \emph{Proc. Interspeech}, 2022, pp. 5513--5517.

\bibitem{yuan2022low}
X.~Yuan, R.~Feng, and M.~Ye, ``{Low-Resource Mongolian Speech Synthesis Based on Automatic Prosody Annotation},'' \emph{arXiv preprint arXiv:2211.09365}, 2022.

\bibitem{rosenberg2010autobi}
A.~Rosenberg, ``{AutoBI - A Tool for Automatic ToBI Annotation},'' in \emph{Proc. Interspeech}, 2010, pp. 146--149.

\bibitem{zou21_interspeech}
Y.~Zou, S.~Liu, X.~Yin, H.~Lin, C.~Wang, H.~Zhang, and Z.~Ma, ``{Fine-Grained Prosody Modeling in Neural Speech Synthesis Using ToBI Representation},'' in \emph{Proc. Interspeech}, 2021, pp. 3146--3150.

\bibitem{stephenson22_interspeech}
B.~Stephenson, L.~Besacier, L.~Girin, and T.~Hueber, ``{BERT, Can HE Predict Contrastive Focus? Predicting and Controlling Prominence in Neural TTS Using a Language Model},'' in \emph{Proc. Interspeech}, 2022, pp. 3383--3387.

\bibitem{sanders2023invert}
N.~Sanders and K.~Richmond, ``{Invert-Classify: Recovering Discrete Prosody Inputs for Text-to-Speech},'' in \emph{2023 IEEE Automatic Speech Recognition and Understanding Workshop (ASRU)}, 2023, pp. 1--7.

\bibitem{suni2017hierarchical}
A.~Suni, J.~{\v{S}}imko, D.~Aalto, and M.~Vainio, ``{Hierarchical Representation and Estimation of Prosody Using Continuous Wavelet Transform},'' \emph{Computer Speech \& Language}, vol.~45, pp. 123--136, 2017.

\bibitem{sloan19_ssw}
R.~Sloan, S.~S. Akhtar, B.~Li, R.~Shrivastava, A.~Gravano, and J.~Hirschberg, ``{Prosody Prediction from Syntactic, Lexical, and Word Embedding Features},'' in \emph{Proc. 10th ISCA Workshop on Speech Synthesis (SSW 10)}, 2019, pp. 269--274.

\bibitem{chen2022character}
X.~Chen, C.~Song, Y.~Zhou, Z.~Wu, C.~Chen, Z.~Wu, and H.~Meng, ``{A Character-Level Span-Based Model for Mandarin Prosodic Structure Prediction},'' in \emph{Proc. ICASSP}, 2022, pp. 7602--7606.

\bibitem{ananthakrishnan2007automatic}
S.~Ananthakrishnan and S.~S. Narayanan, ``{Automatic Prosodic Event Detection Using Acoustic, Lexical, and Syntactic Evidence},'' \emph{IEEE Trans. Audio, Speech, Lang. Process.}, vol.~16, no.~1, pp. 216--228, 2008.

\bibitem{sridhar2008exploiting}
V.~K. Rangarajan~Sridhar, S.~Bangalore, and S.~S. Narayanan, ``{Exploiting Acoustic and Syntactic Features for Automatic Prosody Labeling in a Maximum Entropy Framework},'' \emph{IEEE Trans. Audio, Speech, Lang. Process.}, vol.~16, no.~4, pp. 797--811, 2008.

\bibitem{qian2010automatic}
Y.~Qian, Z.~Wu, X.~Ma, and F.~Soong, ``{Automatic Prosody Prediction and Detection with Conditional Random Field (CRF) models},'' in \emph{2010 7th International Symposium on Chinese Spoken Language Processing}, 2010, pp. 135--138.

\bibitem{gulati2020conformer}
A.~Gulati, J.~Qin, C.-C. Chiu, N.~Parmar, Y.~Zhang, J.~Yu, W.~Han, S.~Wang, Z.~Zhang, Y.~Wu, and R.~Pang, ``{Conformer: Convolution-augmented Transformer for Speech Recognition},'' in \emph{Proc. Interspeech}, 2020, pp. 5036--5040.

\bibitem{devlin2018bert}
J.~Devlin, M.-W. Chang, K.~Lee, and K.~Toutanova, ``{BERT: Pre-training of Deep Bidirectional Transformers for Language Understanding},'' in \emph{Proc. NAACL-HLT}, 2019, pp. 4171--4186.

\bibitem{li2021ppg}
Z.~Li, B.~Tang, X.~Yin, Y.~Wan, L.~Xu, C.~Shen, and Z.~Ma, ``{PPG-Based Singing Voice Conversion with Adversarial Representation Learning},'' in \emph{Proc. ICASSP}, 2021, pp. 7073--7077.

\bibitem{pmlr-v139-radford21a}
A.~Radford, J.~W. Kim, C.~Hallacy, A.~Ramesh, G.~Goh, S.~Agarwal, G.~Sastry, A.~Askell, P.~Mishkin, J.~Clark \emph{et~al.}, ``{Learning Transferable Visual Models from Natural Language Supervision},'' in \emph{Proc. ICML}, 2021, pp. 8748--8763.

\bibitem{wu2023large}
Y.~Wu, K.~Chen, T.~Zhang, Y.~Hui, T.~Berg-Kirkpatrick, and S.~Dubnov, ``{Large-Scale Contrastive Language-Audio Pretraining with Feature Fusion and Keyword-to-Caption Augmentation},'' in \emph{Proc. ICASSP}, 2023, pp. 1--5.

\bibitem{elizalde2023clap}
B.~Elizalde, S.~Deshmukh, M.~A. Ismail, and H.~Wang, ``{CLAP Learning Audio Concepts from Natural Language Supervision},'' in \emph{Proc. ICASSP}, 2023, pp. 1--5.

\bibitem{guo2021recent}
P.~Guo, F.~Boyer, X.~Chang, T.~Hayashi, Y.~Higuchi, H.~Inaguma, N.~Kamo, C.~Li, D.~Garcia-Romero, J.~Shi, J.~Shi, S.~Watanabe, K.~Wei, W.~Zhang, and Y.~Zhang, ``{Recent Developments on Espnet Toolkit Boosted By Conformer},'' in \emph{Proc. ICASSP}, 2021, pp. 5874--5878.

\bibitem{selkirk1980prosodic}
E.~O. Selkirk, ``{On Prosodic Structure and its Relation to Syntactic Structure},'' \emph{Nordic Prosody II}, 1980.

\bibitem{steinhauer2003electrophysiological}
K.~Steinhauer, ``{Electrophysiological Correlates of Prosody and Punctuation},'' \emph{Brain and language}, vol.~86, no.~1, pp. 142--164, 2003.

\bibitem{panayotov2015librispeech}
V.~Panayotov, G.~Chen, D.~Povey, and S.~Khudanpur, ``{Librispeech: An ASR Corpus Based on Public Domain Audio Books},'' in \emph{Proc. ICASSP}, 2015, pp. 5206--5210.

\bibitem{radford2023robust}
A.~Radford, J.~W. Kim, T.~Xu, G.~Brockman, C.~McLeavey, and I.~Sutskever, ``{Robust Speech Recognition via Large-scale Weak Supervision},'' in \emph{Proc. ICML}, 2023, pp. 28\,492--28\,518.

\bibitem{mcauliffe2017montreal}
M.~McAuliffe, M.~Socolof, S.~Mihuc, M.~Wagner, and M.~Sonderegger, ``{Montreal Forced Aligner: Trainable Text-Speech Alignment Using Kaldi},'' in \emph{Proc. Interspeech}, 2017, pp. 498--502.

\bibitem{ljspeech17}
K.~Ito and L.~Johnson, ``{The LJ Speech Dataset},'' \url{https://keithito.com/LJ-Speech-Dataset/}, 2017.

\bibitem{kong2020hifi}
J.~Kong, J.~Kim, and J.~Bae, ``{HiFi-GAN: Generative Adversarial Networks for Efficient and High Fidelity Speech Synthesis},'' \emph{Proc. NeurIPS}, vol.~33, pp. 17\,022--17\,033, 2020.

\end{thebibliography}

\end{document}